\documentclass[journal=jacsat,manuscript=article]{achemso}

\usepackage[version=3]{mhchem} 

\author{Joshua Baxter}
\affiliation{Department of Physics, University of Ottawa}
\alsoaffiliation{Centre for Research in Photonics, University of Ottawa}
\email{jbaxt089@uottawa.ca}
\author{Antonino Cal\`a Lesina}
\affiliation{Department of Physics, University of Ottawa}
\alsoaffiliation{Centre for Research in Photonics, University of Ottawa}
\alsoaffiliation{School of Electrical Engineering and Computer Science, University of Ottawa}
\email{Antonino.Calalesina@uottawa.ca}
\author{Jean-Michel Guay}
\affiliation{Department of Physics, University of Ottawa}
\alsoaffiliation{Centre for Research in Photonics, University of Ottawa}
\alsoaffiliation{Department of Mechanical Engineering, University of Ottawa}
\author{Arnaud Weck}
\affiliation{Department of Physics, University of Ottawa}
\alsoaffiliation{Centre for Research in Photonics, University of Ottawa}
\alsoaffiliation{Department of Mechanical Engineering, University of Ottawa}
\author{Pierre Berini}
\affiliation{Department of Physics, University of Ottawa}
\alsoaffiliation{Centre for Research in Photonics, University of Ottawa}
\alsoaffiliation{School of Electrical Engineering and Computer Science, University of Ottawa}
\author{Lora Ramunno}
\affiliation{Department of Physics, University of Ottawa}
\alsoaffiliation{Centre for Research in Photonics, University of Ottawa}
\title[An \textsf{achemso} demo]
  {Plasmonic colours predicted by deep learning}

\abbreviations{DL,DNN,HL,SEM,FDTD}

\begin{document}


\begin{abstract}
Picosecond laser pulses have been used as a surface colouring technique for noble metals, where the colours result from plasmonic resonances in the metallic nanoparticles created and redeposited on the surface by ablation and deposition processes. This technology provides two datasets which we use to train artificial neural networks, data from the experiment itself (laser parameters vs. colours) and data from the corresponding numerical simulations (geometric parameters vs. colours). We apply deep learning to predict the colour in both cases. We also propose a method for the solution of the inverse problem -- wherein the geometric parameters and the laser parameters are predicted from colour -- using an iterative multivariable inverse design method.  
\end{abstract}

\section{Introduction}
Deep learning (DL) refers to a sub-field of Machine Learning and Artificial Intelligence where multi-layered artificial neural networks, or deep neural networks (DNNs), are used for high accuracy predictions and classifications. Deep learning has been used in speech recognition, image classification, vision, and pattern extraction, to name a few examples \cite{lecun_deep_2015}. Its application in photonics is recent. For example, DL has been used to classify scanning electron microscope (SEM) images of nanostructures \cite{modarres_neural_2017}, to increase the resolution of microscopy \cite{rivenson_deep_2017}, and for the design of nanophotonic structures \cite{ma_deep-learning-enabled_2018,malkiel_plasmonic_2018,pilozzi_machine_2018,peurifoy_nanophotonic_2018,liu_training_2018,sajedian_finding_2018,liu_generative_2018}. DL models can be used to find patterns in complicated datasets and are particularly useful when large pre-calculated or pre-measured datasets are available for training. Both simulations and experiments in nanophotonics can provide the large datasets required for DL. For example, geometric parameter sweeps are used in plasmonics to optimize the design of nanostructures, while parameters are tuned in experiments to produce the best nanofabrication results. The exploration of the parameter space based on trials is time-consuming. Using data from parameter sweeps, one can train a DL model to predict the experimental or simulation outcome for a new set of input parameters, thus avoiding the need to run many trials of the actual experiment or computation.

\begin{figure}[h!]
\centering
\includegraphics[width=6in]{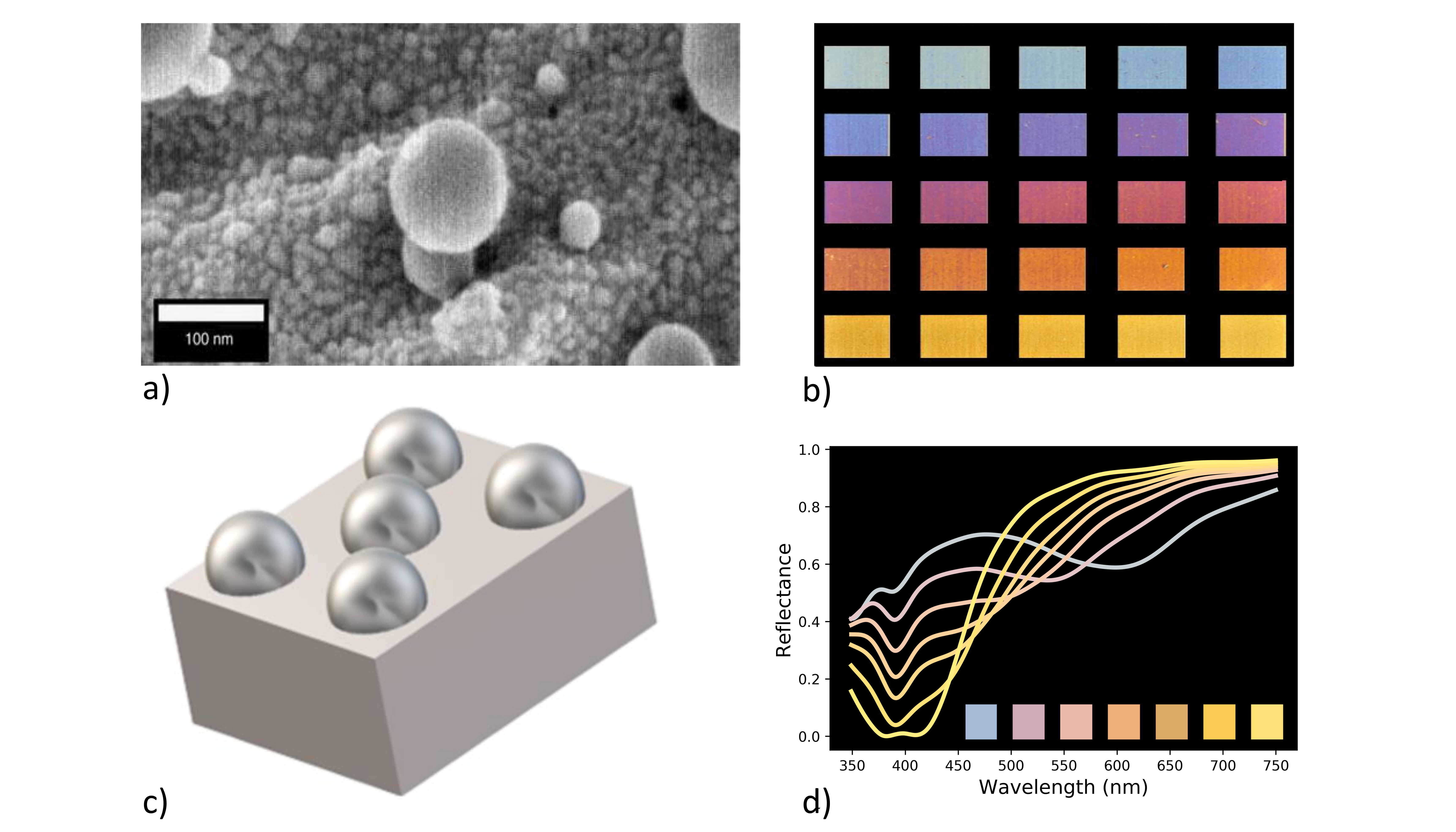} 
\caption{a) High-magnification SEM image of a laser-machined silver surface. b) Colour palette obtained by the laser-machining of silver surfaces . c) Simplified model of the silver surface used in FDTD simulations. d) Simulated reflectance spectra as a function of inter-nanoparticle distance. a) and b) are used and modified (stretched and removed labels from b) with permission from the authors of    \cite{guay_laser-induced_2017} under the Creative Commons Attribution 4.0 International License \cite{ccyb}.}
\label{fig:FirstFig}
\end{figure}

An application which may benefit from DL is the plasmonic colouration of metal surfaces via laser machining. In ref \cite{guay_laser-induced_2017,guay_topography_2018}, by controlling different laser parameters such as scanning speed, hatch spacing, burst number, etc., the authors created laser-machined surfaces (Figure \ref{fig:FirstFig}(a)) with a large range of colours (Figure \ref{fig:FirstFig}(b)). Finite-difference time-domain (FDTD) simulations were performed based on the geometries inspired by the experimental SEM images and the measured nanoparticle distributions, and these simulations were able to explain experimental trends. This includes finding that the colour transition from blue to yellow occurs when the inter-particle spacing for nanoparticles of radius $R_m = 35$ nm is progressively increased. The nanoparticle distribution is modelled as a periodic hexagonal lattice of single-sized nanoparticles embedded into a metal surface (Figure \ref{fig:FirstFig}(c)), and the output of the simulation is a reflection spectrum from which the colour is rendered to the Red-Green-Blue (RGB) colour scale \cite{mathrgb} (Figure \ref{fig:FirstFig}(d)). Both experiments and simulations contain challenges which can be addressed by DL. On the experimental side, the colours generated by a new set of laser parameters must be determined empirically through trial and error. On the computational side, the requirement for accuracy means that a large-scale FDTD simulation has to be run in order to find the colour corresponding to a new set of geometric parameters, and this is time-consuming.

In this paper we apply DL to the direct prediction of new colours using both the experimental and simulation data. We predict colours from laser parameters by leveraging the dataset linking laser parameters to real, physical colours. We also predict colours from nanoparticle geometries by using the dataset linking the simulation geometries to computed colours. We then apply DNNs to aid in the solution of the inverse design problem, whereby inputting colours, the model returns appropriate geometric or laser parameters. In solving the inverse design problem, we use an iterative multivariable inverse design method in which the inverse solution is found by iteratively solving for individual outputs. To the best of our knowledge, this is the first time this method has been reported. Traditional optimization techniques (e.g. genetic algorithms) would be time consuming for the large-scale simulations, and costly to implement in the lab. This is why we highlight the importance of using the data that has already been accumulated to train efficient DL models.

\section{Direct Prediction of Colours}

\subsection{Using Simulation Data}

\begin{figure}[h!]
\centering
\includegraphics[width=3.33in]{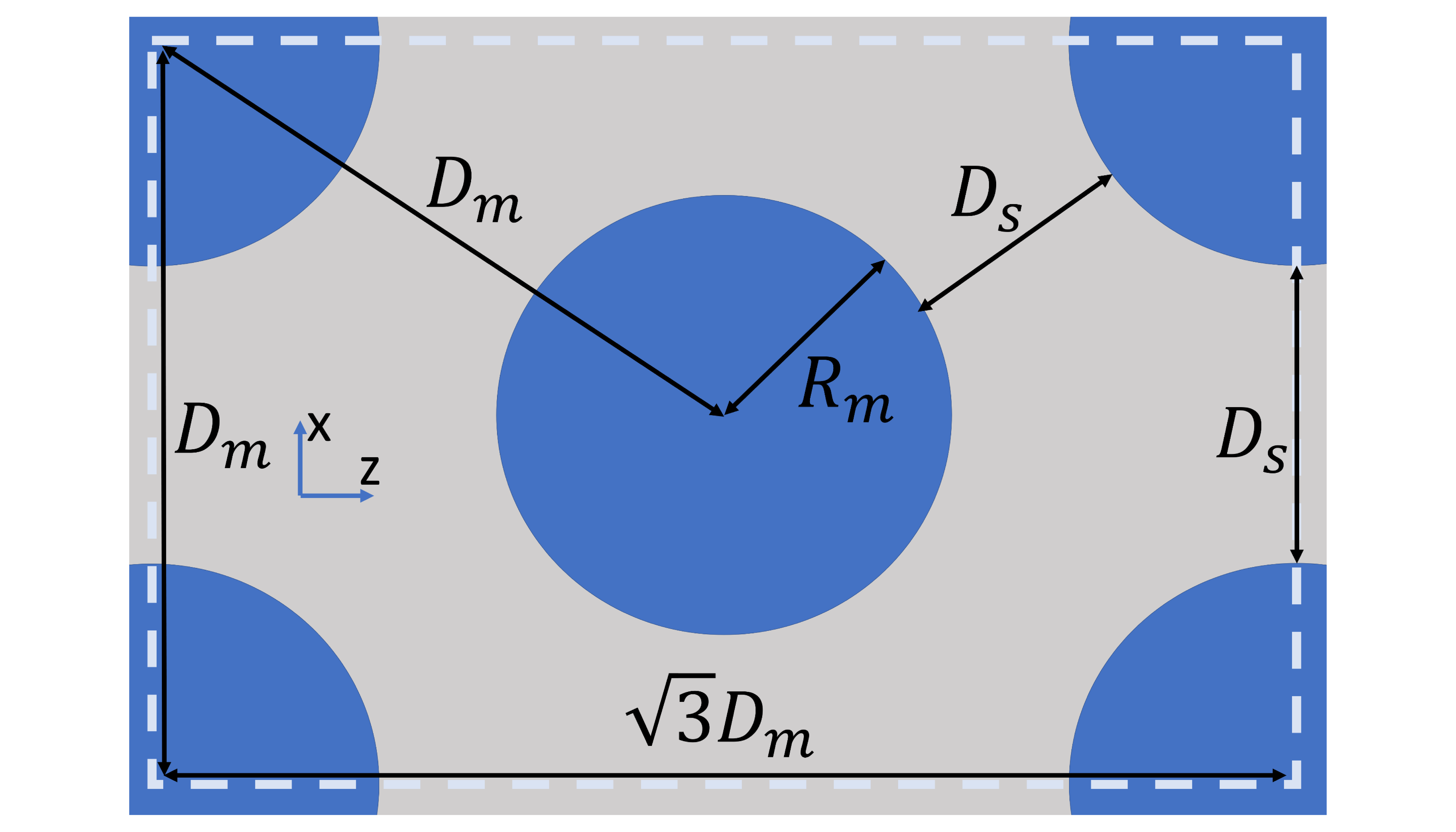}
\caption{Layout of the nanoparticle geometry in the x-z plane. The white dashed box is the unit cell used in the FDTD simulations}
\label{fig:FDTDLayout}
\end{figure}

In this section we use DNNs for the prediction of colours from nanoparticle geometries using the FDTD dataset. The FDTD simulations were run to calculate the reflectance spectra from silver nanoparticle distributions embedded into flat silver surfaces. The particles are arranged in a hexagonal lattice in the x-z plane as shown in Figure \ref{fig:FDTDLayout}, where only a unit cell is simulated and periodic boundaries are placed in the x- and z- directions. The incoming broadband pulse is propagated in the y-direction which is truncated by Convolutional Perfectly Matched Layers (CPMLs). Simulations are run with particle radii $R_m$ ranging from 5 nm to 100 nm. The embedding depth represents the percentage of the radius embedded into the silver surface, and it ranges from 10 to 100\%, where an embedding of 100\% means that we simulate a hemisphere. The inter-particle spacing $D_m$ describes the center-to-center spacing between the nanoparticles. To remove correlation between spacing and radius, we also define an edge-to-edge spacing $D_s  = D_m  - 2R_m$ which ranges from 5 nm to 80 nm. The simulations used a space-step of $dx$ = 0.1 nm for the $R_m$= 5 nm particles, $dx$ = 0.25 nm for the $R_m$= 10 to 20 nm particles and $dx$ = 0.5 nm for the $R_m$= 25 to 100 nm particles. The optical parameters of silver were modelled by the Drude + 2 Critical Points model \cite{prokopidis_unified_2013}. The total number of Yee cells \cite{yee_numerical_1966} depends on the space step and geometry but is on the order of $10^8$. The length of the y-dimension is taken to guarantee that the reflectance is computed at a distance from the surface located in the far-field region. The reflectance spectra were computed over the optical wavelength range of 350-750 nm and were converted to RGB values. The simulations used an in-house 3D-FDTD parallel code\cite{lesina_convergence_2015} running on the IBM BlueGene/Q supercomputer (64k cores)\cite{BGQRef} operated by SOSCIP at the University of Toronto, Canada.

For the direct prediction DL application, the input is the simulation geometries which are parameterized by nanoparticle radius $R_m$, embedding depth, and edge-to-edge nanoparticle spacing $D_s$. The output is one of the colour values R, G, or B. Our total simulation dataset consists of 770 datapoints.

To optimize the use of DL, we had to find the most ideal DNN architecture: the number hidden layers (HLs) and the number of nodes in each HL that best models our dataset. N-fold Cross Validation (CV)\cite{james_introduction_2013} is a process where the data is separated into N subsets. A DNN is trained using N-1 of the subsets. The Nth subset is used for testing and the error is recorded. This process is repeated N times so that each subset is used for testing. The error of a given DNN architecture is the average error from the N DNNs on the test sets.

Ten-fold CV is applied to DNNs with HL number ranging from 1 to 10, and nodes per HL ranging from 4 to 72. We simplify this process by only making DNNs with equal numbers of nodes per HL. Seeing as each DNN outputs only one value (be it R, G, or B), this process is repeated for each colour component. To find an ideal DNN architecture we decided to find one with a minimum sum of normalized Mean Square Errors (MSEs):
\begin{equation}
min\Bigg(\frac{R_{MSE}}{{min(R_{MSE})}} + \frac{B_{MSE}}{{min(B_{MSE})}} + \frac{G_{MSE}}{{min(G_{MSE})}}\bigg)
\end{equation}
where $R_{MSE}$, $G_{MSE}$, and $B_{MSE}$ are the mean squared error for the R, G and B values respectively. We also want a DNN that uses a minimal number of HLs and nodes, in order to avoid over-fitting and long training times.

Because of the large number of DNNs trained in the CV process, we made use of the SOSCIP GPU cluster \cite{GPURef} where we could train 4 DNNs consecutively on a single node consisting of 4 GPUs. The DNN codes were written using TensorFlow\cite{TensorFlowRef} and the weights of the DNNs were optimized using an Adam Optimizer\cite{kingma_adam:_2014}.

\begin{figure}[h!]
\centering
\includegraphics[width=3.33in]{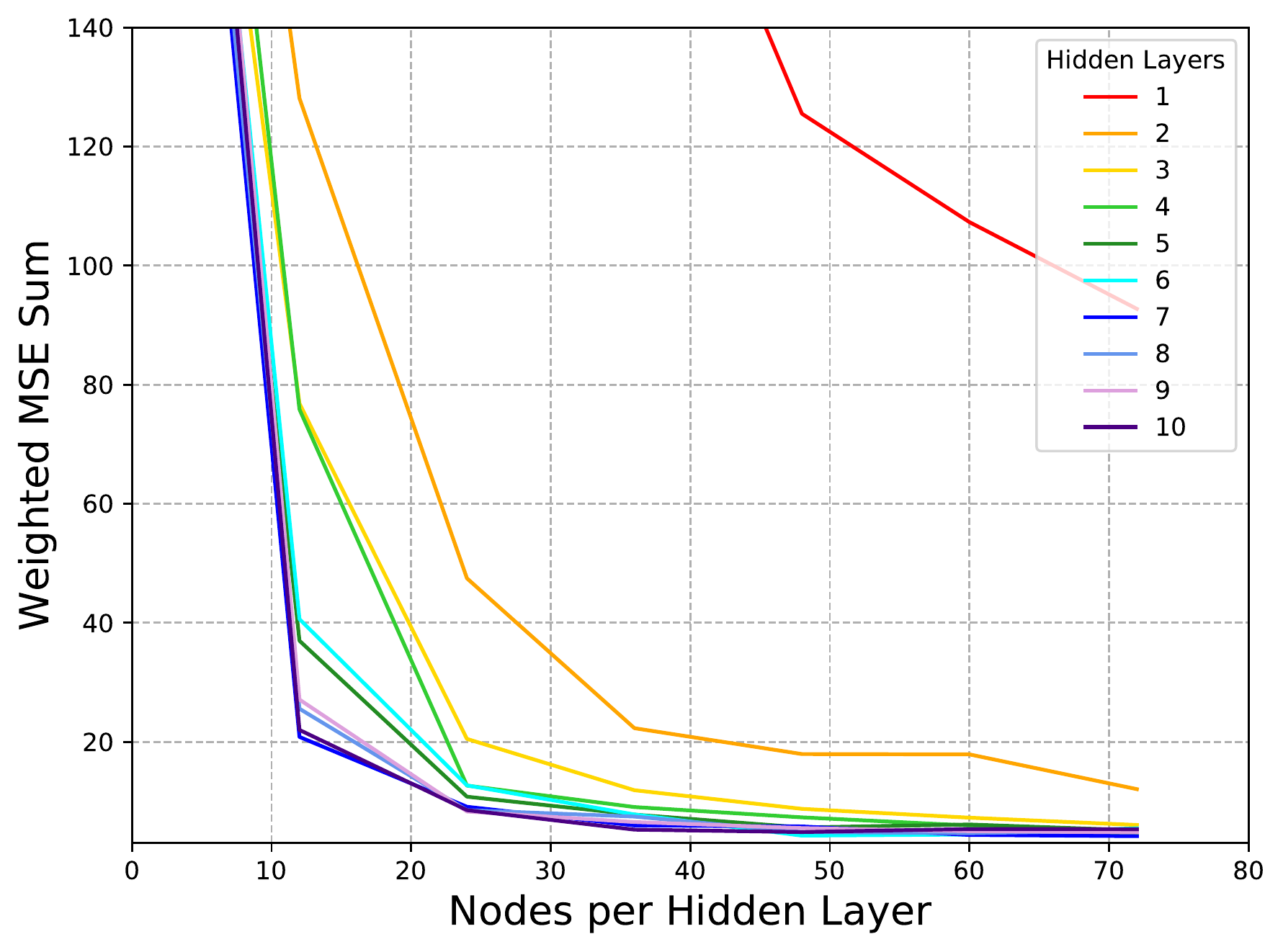}
\caption{Weighted sum of the MSE for different DNN architectures for the simulation dataset}
\label{fig:MSEError}
\end{figure}

\begin{figure}[h!]
\centering
\includegraphics[width=6in]{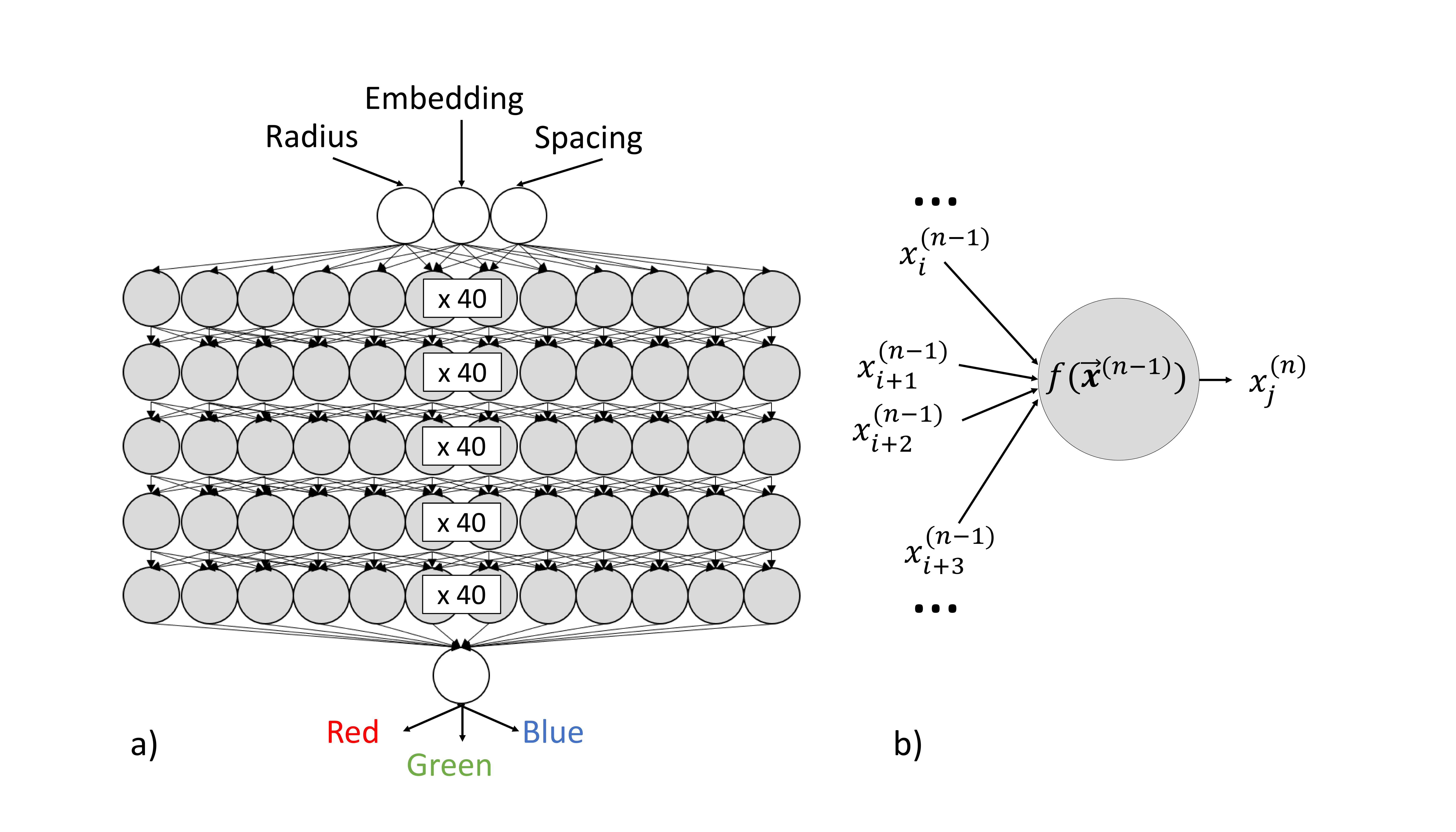}
\caption{a) Sketch of the DNN used on the simulation dataset. b) The $j^{th}$ hidden layer node of the $n^{th}$ hidden layer.}
\label{fig:BasicDNN}
\end{figure}

Figure \ref{fig:MSEError} shows the normalized MSE sum as a function of nodes per HL for all of HLs considered. We see that the error plateaus for five or more HLs, after 40 nodes per HL. Beyond this point, the DNNs all have similar accuracies. Because of this, five HL DNN with 40 nodes per HL, as sketched in Figure \ref{fig:BasicDNN}(a), was chosen to model the simulation dataset.

A node in a hidden layer will receive as inputs the outputs from a previous layer, as shown in Figure \ref{fig:BasicDNN}(b). The node performs a linear transformation on the inputs:
\begin{equation}
    f(\mathbf{x}^{(n-1)}) = b_j^{(n)} + \sum_i a_{i,j}^{(n)} x_i^{(n-1)}
\end{equation}
where $a_{i,j}^{(n)}$ is the weight of the $j^{th}$ node in the $n^{th}$ hidden layer applied to the $i^{th}$ input, and $b_j^{(n)}$ is the bias of the $j^{th}$ node in the $n^{th}$ hidden layer. The output of this node is given by the rectified linear activation function:

\begin{equation}
    x_j^{(n)} = \textup{max}(f(\mathbf{x}^{(n-1)}),0)
\end{equation}

The training process involves adjusting the weights $a_{i,j}^{(n)}$ and the biases $b_j^{(n)}$ to minimize the error on the training set. This DNN was trained using 90\% of the available data. The remaining randomly selected 10\% not used in the training process was used for testing. 

\begin{figure}[h!]
\centering
\includegraphics[width=6in]{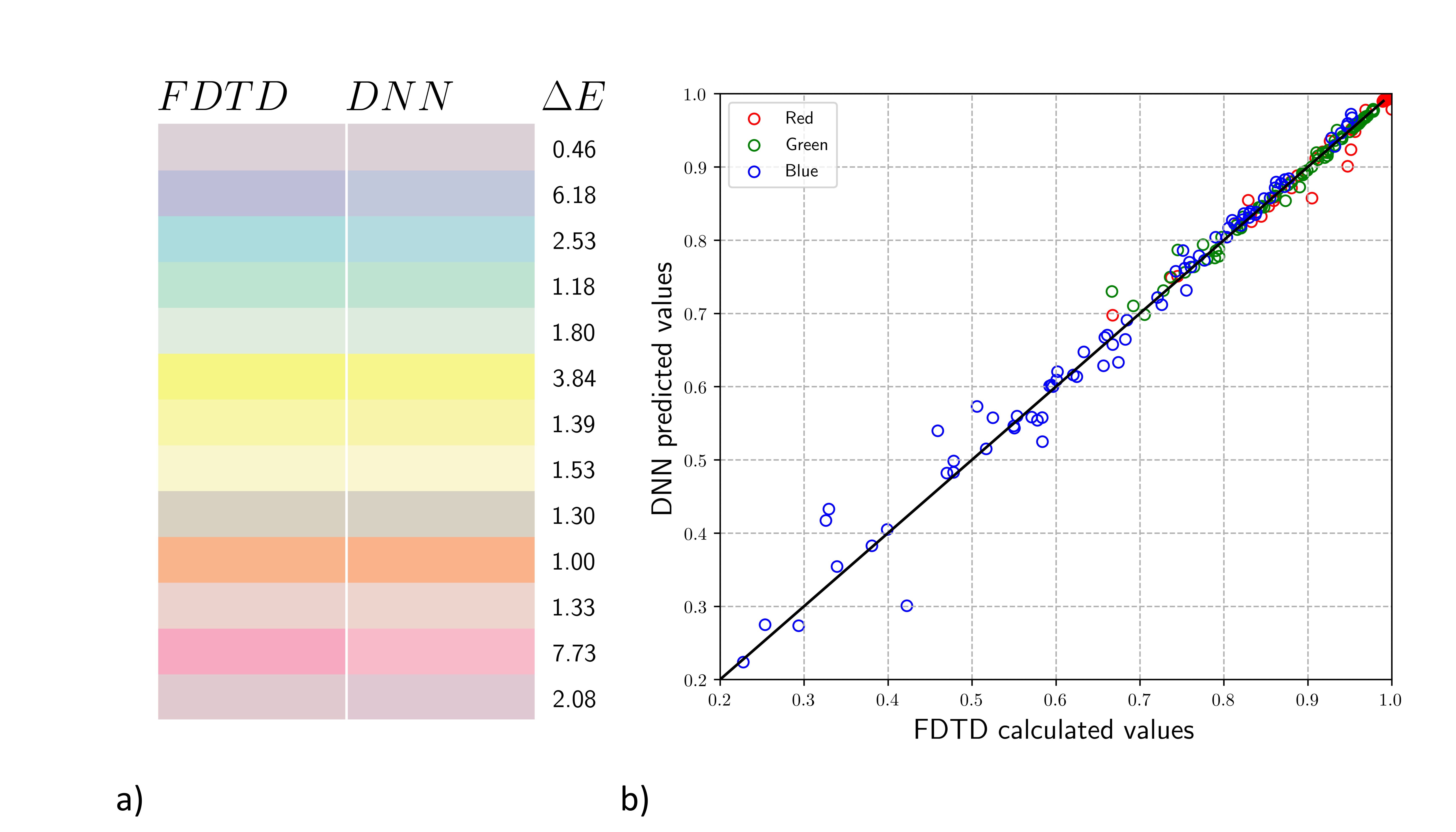}
\caption{FDTD simulation colours vs. DNN prediction. a) Comparison between FDTD simulation colours (left) and DNN predicted colours (right) on the test set with corresponding $\Delta E$ values.  b) Comparison between RGB values from the simulation results (x-axis) and the DNN predictions (y-axis) along with the ideal linear model (y = x). The colour of each circle represents whether the point represents an R (red), G (green), or B (blue) datapoint.}
\label{fig:SimDirect}
\end{figure}

Figure \ref{fig:SimDirect}(a) shows a palette of some of the colours used in the test set, where the calculated colours from the FDTD simulations are shown on the left and the DNN predicted colours are shown on the right for the same geometric parameters. To the right of the DNN colour blocks we give the $\Delta E$ value \cite{guay_passivation_2018} which is a quantitative measure of the difference between two colours described by: 
\begin{equation}
    \Delta E = \sqrt{(L_1 - L_2)^2 + (a_1 - a_2)^2 + (b_1 - b_2)^2}
\end{equation}
where $L_i$, $a_i$, and $b_i$ (i=1,2)  are the colour coordinates in an alternate colour space, calculated from the RGB values\cite{mathrgb}. For reference, a $\Delta E$ of 1 is the threshold of a colour difference that is just perceptible to the human eye, and 6-7 (as a subjective threshold) is often deemed acceptable in commercial practice. In the test set of 77 colours, only 8 surpass a $\Delta E$ of 7. For example, the penultimate block in Figure \ref{fig:SimDirect}(a) yields $\Delta E$=7.73, yet the colour difference is only somewhat noticeable. The mean $\Delta E$ of the entire test set is 2.61. The colour palette and corresponding $\Delta E$ demonstrate the predictive accuracy of our model. 

Figure \ref{fig:SimDirect}(b) compares the RGB values of FDTD vs DNN for the entire test set.  Each circle represents a single datapoint, where the x-axis is the R, G, or B value of the FDTD calculation, and the y-axis is the corresponding R, G, or B value predicted by the DNN. The colour of the circle is either red, green or blue, corresponding to which RGB component that datapoint represents. The ideal linear model (y = x) is also displayed for convenience as a solid line. The mean percent error on the test set for R, G, and B are 0.48\%, 0.67\%, and 3.31\% respectively, showing that the DNN chosen predicts all three value with very good accuracy. The larger error for the blue is likely due to the relative lack of representation of low blue values in the dataset. The DNN was not given enough training data with blue values below 0.6 where we clearly see the most error in Figure \ref{fig:SimDirect}(b). We can see from the plot that numerical values are well predicted from the model.

It is worth noting that this model can only be used for interpolation and very small extrapolations. Large extrapolations can lead to nonphysical results. Over the range spanned by our data, this model can predict the colours with high accuracy (less than 4\% error) without having to run additional simulations.

\subsection{Using Laser Machining Data}

In this section we repeat the above analysis for the laser parameter dataset. This dataset consisted originally of 10 laser parameters and 1841 datapoints. We eliminated 6 laser parameters due to lack of variation and effect on colour, which left 4 laser parameters under consideration: laser scanning speed, hatch spacing, number of laser pulses per burst, and laser fluence. Physically these parameters affect the total accumulated fluence on the metal surface. This was found empirically to be the key factor in colour production, so that different laser parameters with the same total accumulated fluence produced very similar colours\cite{guay_laser-induced_2017}. While eliminating laser parameters, we also removed the corresponding datapoints to ensure that the training process was not affected. The number of datapoints in our updated set is 1745.

We use the process outlined in the previous section to find the ideal DNN for this dataset. We choose our model to be a DNN with 3 HLs and 60 nodes per hidden layer.  

\begin{figure}[h!]
\centering
\includegraphics[width=6in]{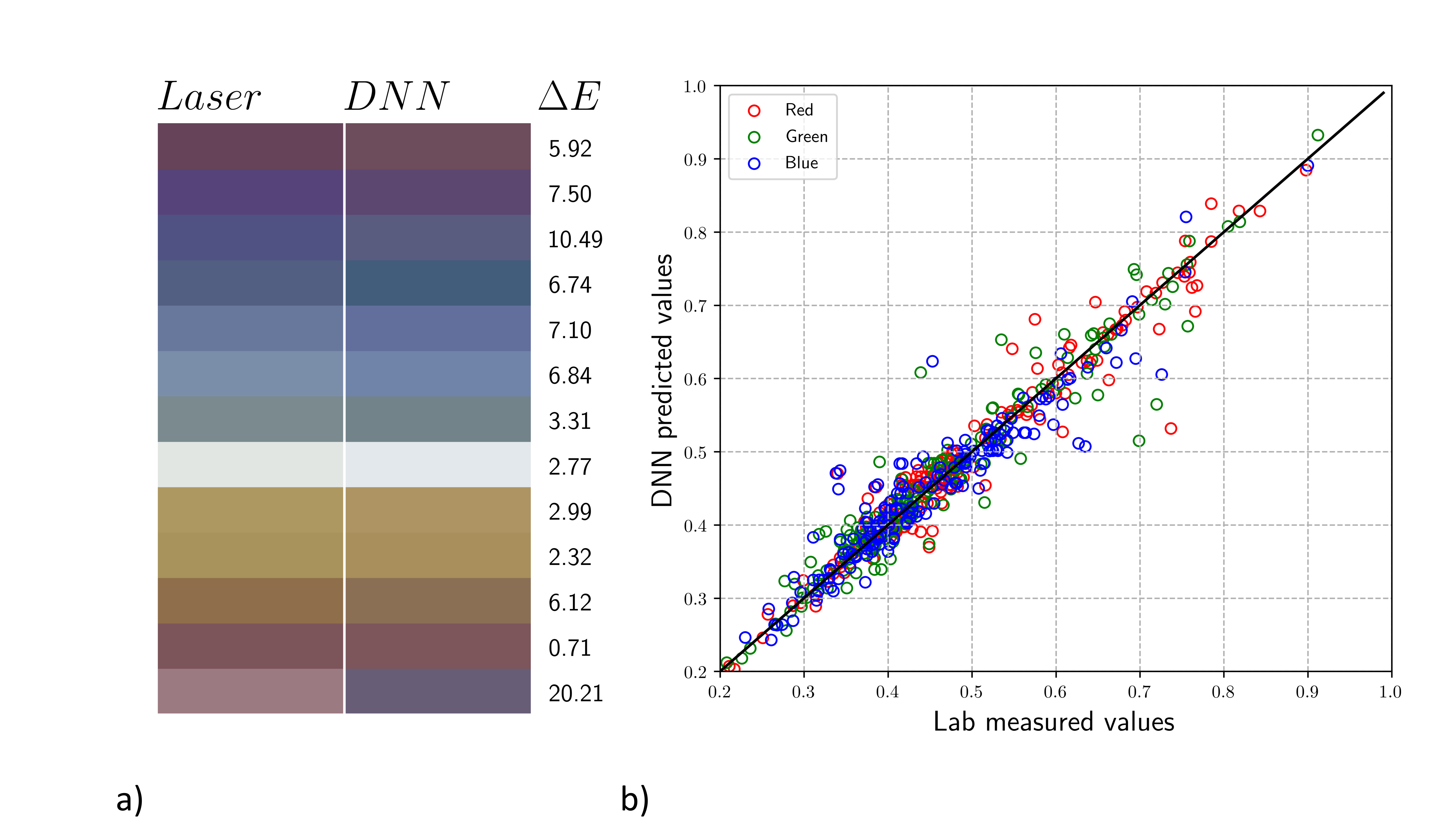}
\caption{Laser parameter colours vs. DNN prediction. a) Comparison between the measured colours (left) and DNN predicted colours (right) with corresponding $\Delta E$ values. b) Comparison between RGB values from the experimental measurements (x-axis) and the DNN predictions (y-axis) along with the ideal linear model (y = x). The colour of each circle represents whether the point represents an R (red), G (green), or B (blue) datapoint.}
\label{fig:ExpDirect}
\end{figure}

We split our data randomly into two sets: 90\% for training and 10\% for testing. We again display a comparison between some of the DNN predicted colours and those measured in the lab in Figure \ref{fig:ExpDirect}(a) along with the $\Delta E$ for the test set. The error is much more significant for this dataset in comparison to the simulation dataset. There is a much more complicated relationship between laser parameters and the RGB of the resulting colour. Despite this, the DNN has found enough relationships to be able to predict a colour within a given range of the measured colours, resulting in a mean $\Delta E$ of 5.27 for the test set, where 41 of the 145 test datapoints are greater than 7.  We can see a few of these in Figure \ref{fig:ExpDirect}(a), as the second, third, fifth, and last blocks from the top. The first and third, although having high $\Delta E$ values, are still subjectively well predicted. The last block however is the result of a large difference (greater than 0.1) in the R, G and B values. Large errors like these are rare but should be expected in a large test set such as this. In a large test set, there will likely be points that are not well represented in the training data. This means that the DNN is not as well trained in these regions of parameter space. 
In Figure \ref{fig:ExpDirect}(b) we display the comparison of RGB values for the entire test set. The mean percent error for R, G and B are 4.29\%, 4.88\%, and 5.47\% respectively. We see that this DNN predicts each colour reasonably well. However we point out that care should be taken when applying DL to the prediction of subjective results – here, even a small error can change the colour perceptively. 

\section{Inverse Design Prediction from Colours}
\subsection{Using Simulation Data}

In this section we solve the inverse problem: we find the simulation geometries that are required to produce given colours, where the RGB values are now the inputs. Solving the inverse problem is not a trivial task as the solution may not be unique. Multiple geometries may produce the same colour so a DNN cannot be directly trained to predict geometric parameters from a given colour. Methods to get around the uniqueness problem have been suggested, including (1) training the inputs using back-propagation with the weights of the DNN fixed\cite{peurifoy_nanophotonic_2018}, (2) training the inverse DNN using the forward DNN\cite{liu_training_2018,liu_generative_2018}, and (3) applying reinforcement learning\cite{sajedian_finding_2018}. Here we introduce a more straightforward iterative multivariable inverse design approach, which to our knowledge remains unreported. 

Our approach involves training $n$ neural networks; \{$DNN_1$,$DNN_2$,...,$DNN_n$\}, where $n$ is the number of outputs (in this case our geometric parameters), given by \{$O_1$,$O_2$,...,$O_n$\}. $DNN_1$ is trained to calculate output $O_1$ from the desired inputs (in this case the RGB values) and the remainder of the outputs \{$O_2$,...,$O_n$\}. This means that the remainder of the outputs act as inputs for $DNN_1$ alongside the desired RGB values. $DNN_2$ is trained to calculate $O_2$ using the desired inputs and the remainder of the outputs \{$O_1$,$O_3$,...,$O_n$\}, and so on. 

Once all $n$ DNNs are trained, we can calculate the set of outputs in an iterative procedure. We first calculate $O_n$ using the $DNN_n$ with the desired inputs and the remaining outputs (which can be initialized as random numbers or the mean values of the dataset). Next we calculate $O_{n-1}$ using $DNN_{n-1}$ with the desired inputs and the remaining outputs including the updated $O_n$. This is repeated for all outputs. This process can be repeated until the values of the outputs have relaxed to a final value. This method is very simple in implementation since it requires no special treatment of the DNNs unlike inverse design methods suggested in previous studies \cite{peurifoy_nanophotonic_2018,liu_training_2018,liu_generative_2018,sajedian_finding_2018}.   

\begin{figure}[h!]
\centering
\includegraphics[width=6in]{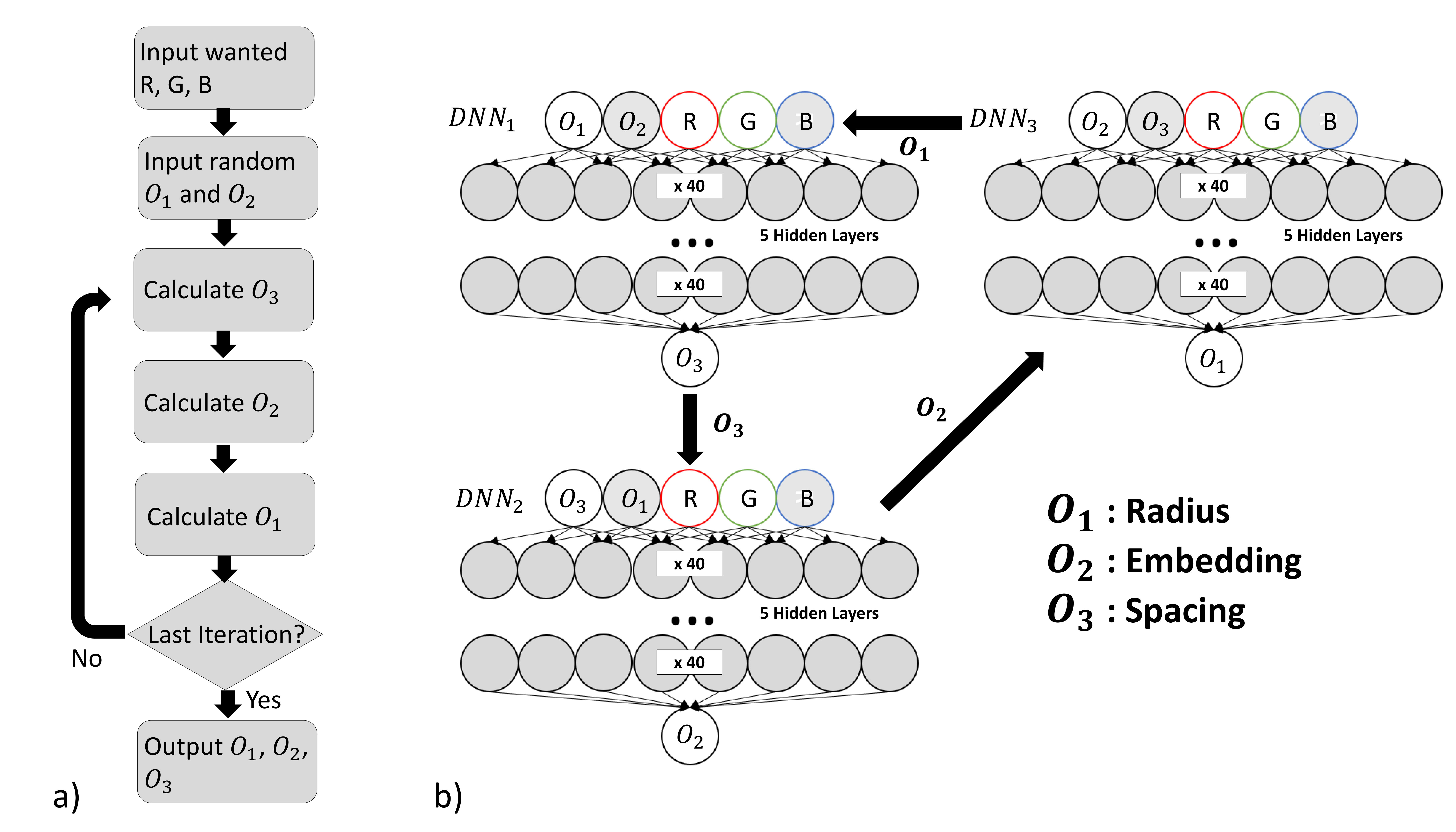}
\caption{a) Iterative multivariable inverse design algorithm for the prediction of geometric parameters from colours. We input the desired R, G, B and initial (random or mean) values for the radius and embedding. We then iterate through DNN calculations for spacing, embedding, radius until the last iteration. b) DNN calculation iteration procedure.}
\label{fig:InvMethod}
\end{figure}

In the case of our simulation geometry, $n$ = 3. $DNN_1$ is trained to predict the particle spacing from R, G, B, radius, and embedding. $DNN_2$ predicts the embedding from R, G, B, radius, and spacing. $DNN_3$ predicts the radius from R, G, B, embedding, and spacing. After training these three DNNs, we can then calculate the three geometric parameters using the iterative procedure outlined in Figure \ref{fig:InvMethod}(a).

First the radius and embedding are set as the mean radius and mean embedding of the dataset, providing a good starting place for the parameter search. We then calculate the spacing using our spacing DNN. The calculated spacing value and mean radius value are used to predict the embedding. This embedding and the spacing are then used to calculate the radius. We then repeat these steps until the geometric parameters relax to their final values. This generally this takes 10-20 iterations, however, this may vary depending on the colours themselves. If the colours are not well represented in the training data, the number of iterations will increase. As sketched in Figure \ref{fig:InvMethod}(b), we use 5 HLs with 40 nodes per hidden layer for our backwards DNNs. 

\begin{figure}[h!]
\centering
\includegraphics[width=6in]{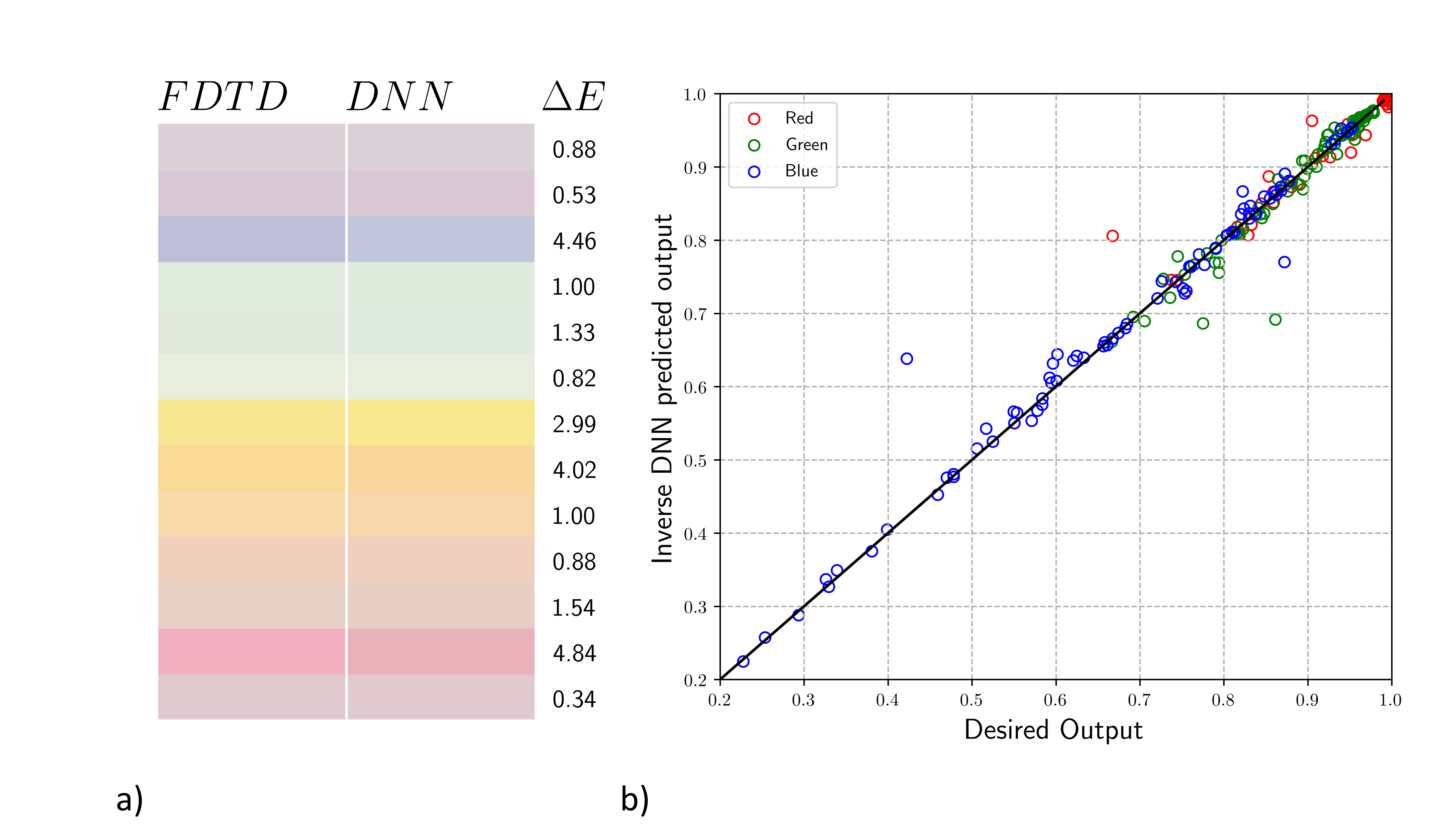}
\caption{Geometric parameter inverse DNN prediction. a) Comparison between the input test colours (left) and the colours produced by the output geometric parameters (right) with corresponding $\Delta E$ values. b) Comparison between input RGB values (x-axis) and the values produced by the output geometric parameters (y-axis) along with the ideal linear model (y = x). The colour of each circle represents whether the point represents an R (red), G (green), or B (blue) datapoint.}
\label{fig:SimInv}
\end{figure}

A sample colour palette of our test dataset is shown in Figure \ref{fig:SimInv}(a) where on the left we display a set of desired colours, and on the right we display the colours generated by the output geometric parameters. We see an agreement between the input colour, and that generated by our iterative method. We also see $\Delta E$ values within the same range as the direct prediction with a mean of 2.83. The comparison between input and output RGB values are given in Figure \ref{fig:SimInv}(b) where the percent error between the input and output colours are 0.83\%, 1.34\%, and 2.12\% for R, G and B respectively. The slight increase in error between the forward and inverse prediction is expected due to error accumulation between the backward DNNs used to calculate the geometries and the forward DNN used for testing the geometries. The geometries output by the relaxation are not guaranteed to provide the closest possible colour to the input. They depend on the initial inputs ($O_1$ and $O_2$ in Figure \ref{fig:InvMethod}). Multiple initialized inputs can be used to find multiple possible geometries, and the one which has the least error between input and output colours should be used.

\subsection{Using Laser Machining Data}

The laser parameter dataset was obtained from the Royal Canadian Mint, who uses this colouring technique to create artwork on silver\cite{guay_laser-induced_2017}. The colours for this artwork are chosen carefully using a palette of known experimentally-obtained colours. The laser is then programmed with the corresponding laser parameters to create this colour on the metal surface. This process is restricted to the list of known colours. However, by using DNNs to predict the laser parameters required to produce a given desired colour, this process can be simplified and laser parameters can be predicted for colours that are not within the known list. 

\begin{figure}[h!]
\centering
\includegraphics[width=6in]{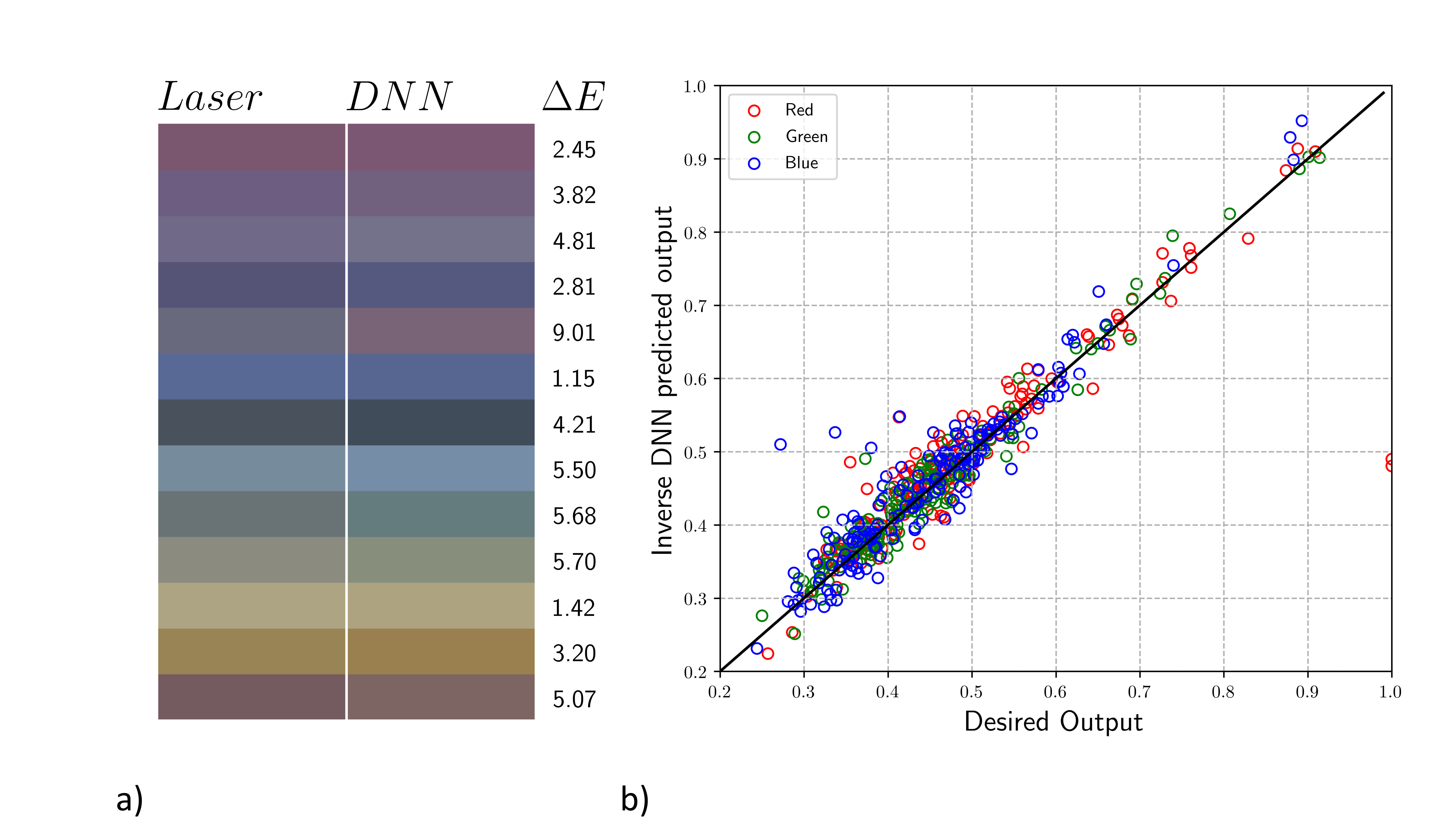}
\caption{Laser parameter inverse DNN prediction. a) Comparison between the input test colours (left) and the colours produced by the output laser parameters (right) with corresponding $\Delta E$ values. b) Comparison between input RGB values (x-axis) and the values produced by the output laser parameters (y-axis) along with the ideal linear model (y = x). The colour of each circle represents whether the point represents an R (red), G (green), or B (blue) datapoint.}
\label{fig:ExpInv}
\end{figure}

We use the iterative multivariable inverse design procedure introduced in the previous section with $n$ = 4 for the four laser parameters that we want to predict (the same ones as used for Figure \ref{fig:ExpDirect}). We train our four backward DNNs with 90\% of the data and test it with the other 10\%. These DNNs have three HLs and 60 nodes per HL. We use the direct prediction DNN to calculate the colour from the predicted laser parameters. We show some of the colours output in Figure \ref{fig:ExpInv}(a) and display the numerical results in Figure \ref{fig:ExpInv}(b). The percent error between the input and output colours are 5.50\%, 4.57\%, and 6.09\% for R, G, and B, respectively, and the mean of $\Delta E$ is 6.31, which remains less than our limit of 7. Numerically these errors are small and result in the linear pattern displayed in Figure \ref{fig:ExpInv}(b). However, these errors are large enough that some colours are switched in the process. For example, the fifth box from the top in the palette of Figure \ref{fig:ExpInv}(a) reveals a change from the desired grey to the output purple. Such changes are caused by differences in the RGB values of the order of 0.07. These differences should be expected due to the accumulation of error in experiment, the backward DNNs, and the forward DNN used to test the backward DNNs. Despite the error in some of the colours, the majority of them are accurately predicted.

\section{Conclusion}
We have applied deep learning to plasmonic metal colouration through direct prediction of colours and by introducing a relaxation technique for the inverse design problem. We found reasonable prediction accuracy for colours produced by deterministic nanoparticle distributions and by surfaces produced by picosecond laser irradiation. The iterative multivariable inverse design method allows for the extraction of geometric parameters and laser parameters from input colours, and this too was demonstrated with reasonable accuracy. The direct prediction and inverse design method presented in this paper have useful applications. The direct prediction of colours from new parameters means that we can fill in blank spaces in our data without needing to run further experiments. Such interpolation can be used to further expand the dataset. Small extrapolations can also be used to expand the dataset however this should be done with care. The ability of deep neural networks to find patterns in complicated data can also be exploited for physical interpretation of the data, however that is beyond the scope of this paper. The inverse design method is applicable especially in the experimental context. The Royal Canadian Mint has started using laser colouring on silver\cite{guay_laser-induced_2017} where the appropriate colours are chosen manually from the database. The inverse design algorithm can automate this process by finding the laser parameters (that are not necessarily in the database) that best match the desired colour assuming the colour is one that can be created by the laser. The applications of inverse design are numerous in plasmonics and nanophotonics\cite{molesky_inverse_2018} and applying advanced methods such as genetic algorithms and reinforcement learning can be daunting. The iterative inverse design procedure introduced here is simple to implement for those who may be new to the field of deep learning provided they have the data.

\begin{acknowledgement}

SOSCIP is funded by the Federal Economic Development Agency of Southern Ontario, the Province of Ontario, IBM Canada Ldt., Ontario Centres of Excellence, Mitacs and Ontario academic member institutions. The authors thank SOSCIP for their computational resources and financial support. We acknowledge the computational resources and support from Scinet. We acknowledge financial support from the National Sciences and Engineering Research Council of Canada, and the Canada Research Chairs program. The authors also thank the Royal Canadian Mint for the use of their laser lab and the data acquired from it. We would like to thank Graham Killaire, Meagan Ginn,  Guillaume C\^ot\'e, and Martin Charron for their contributions in creating the colour palettes at the Royal Canadian Mint. 

\end{acknowledgement}





\bibliography{Plasmonic}


\end{document}